# Towards real-time population estimates: introducing Twitter daily estimates of residents and non-residents at the county level

Yago Martín[a*], Zhenlong Li[b], Yue Ge [a]

[a]*School of Public Administration, University of Central Florida, Orlando, United States;*
[b]*Geoinformation and Big Data Research Laboratory, Department of Geography, University of South Carolina, Columbia, United States,*

[*]Corresponding author: ym@ucf.edu

**Abstract**

The study of migrations and mobility has historically been severely limited by the absence of reliable data or the temporal sparsity of the available data. Using geospatial digital trace data, the study of population movements can be much more precisely and dynamically measured. Our research seeks to develop a near real-time (one-day lag) Twitter census that gives a more temporally granular picture of local and non-local population at the county level. Leveraging geotagged tweets to determine the home location of all active Twitter users, we contribute to the field of digital and computational demography by obtaining accurate daily Twitter population stocks (residents and non-residents). Internal validation reveals over 80% of accuracy when compared with users self-reported home location. External validation results suggest these stocks correlate with available statistics of residents/non-residents at the county level and can accurately reflect regular (seasonal tourism) and non-regular events such as the Great American Solar Eclipse of 2017. The findings demonstrate that Twitter holds potential to introduce the dynamic component often lacking in population estimates.

## 1. Introduction

The spatial mobility capacity of humans has progressed significantly throughout history. Not far ago, people were only able to use their own power to move across space. In some instances, they were able to accelerate the travel speed aided by animal power such as horses or propelled by the force of the wind. The capacity to move found a breakpoint with the arrival of the rail transportation, but it was not until the revolution of transportation in the 20$^{th}$ century when mobility became accessible to the great majority of the population. The development of the automobile and the road system, as well as the arrival of air transportation, completely transformed human mobility. Mobility transitioned from the rare privilege of a few to the daily reality of the many. This trend continues to be accelerated, and new advances in the transportation sector promise to increase the speed and commodity of travel, making spatial mobility even more prevalent than today. In addition, external factors such as weather-related hazards and climate change are thought to have the potential to increase the migration response in the following decades (IPCC, 2014).

Spatial mobility has long been a matter of interest for many disciplines (e.g., Faggian and McCann, 2008; Squire, 2010; Colleoni, 2016), and particularly as a subject of study of geography (Cresswell, 2011). In a context of unprecedented human movement, where short and long-distance travel becomes more and more common, the study of all kinds of mobility is a necessity. The study of population movements has historically been severely limited by the absence of reliable data or the temporal sparsity of the available data (Laczko, 2015; Rango and Vespe, 2017), and it will continue to be capped if the research community does not find the appropriate mechanisms to track individuals over time and across space. For long-term movements (e.g., displacement, migration), census data, with a multi-year periodicity, and household surveys, often collected on an annual basis, constitute the main data source for studying population movements. Address registers, residence permits, and visas record have also been analyzed by researchers when available. However, in broad terms, changes of residence are rarely recorded and readily available at the time of the move, which precludes from connecting triggering events with population movements (Fussell et al. 2014). Shorter-term movements (e.g., commuting, vocational movements, or evacuations) need to rely on individual surveys or transportation statistics, which in many cases do not provide comparable information. The need for a comparable, reliable, and dynamic (with a finer temporal resolution) source of data is, therefore, a condition to improve our understanding of human spatial mobility.

In addition to the transportation revolution, the advancements in electronics and computation have permitted to reach a high degree of physical and virtual connectivity that Harvey (1989) named time-space compression. The digital revolution meant a shift from analog electronic technology to digital electronics, particularly after the propagation of the Internet and the mass production of computers and cell phones.. In this digital era, the immense amount of data produced by the use of digital devices and web-based platforms is commonly referred to as "Big Data", which many characterized as the "4+1 Vs": Volume, Velocity, Variety, Veracity, and more recently, Value (Laczko, 2015). Several researchers believe that Big Data holds the capacity to transform the study of spatial mobility (Isaacson and Shoval, 2006; Billari and Zagheni, 2017; Cesare et al., 2018). Within the great variety of Big Data types, the geospatial digital trace data, also known as passive citizen sensor data, is seen as the source of data that can capitalize this change. However, the research community, with very limited access to these geospatial digital trace data, has not yet fully harness its potential and has not been able to produce population statistics with more suitable spatiotemporal resolutions for the study of population dynamics.

Our research seeks to develop a near real-time (one-day lag) Twitter census that gives a more temporally granular picture of local and non-local population at the county level. Using geotagged tweets to determine the home location of all active Twitter users, we aim to make a substantial contribution to the field of digital and computational demography and obtain daily estimates of Twitter population stocks (residents and non-residents) for the state of South Carolina (SC), as a first step for extending this analysis to the whole United States (US) and other parts of the world. We believe the resulting output can fuel numerous research projects in several fields such as disaster management and public health and help reach a better understanding of human mobility. For instance, the creation of a world-wide real-time monitoring of the spatial behavior of the population would be particularly relevant during emergencies such as the COVID-19 crisis, where identifying where population are concentrating in real time or determining whether people are complying with shelter-in-place official orders becomes a first-class necessity for authorities. Also, this systematic data collection and processing would allow a comparative measure of the impact of a shock such as COVID-19 on the tourism and help tracking the recovery process of the sector. In this sense, several recent publications have attempted to track the spread of the virus using geotagged tweets (Bisanzio et al. 2020) or measure the mobility response during the pandemic (Huang et al. 2020).

The paper is organized as follows. First, we review the most common demographic approaches to the study of short-term and long-term population movements followed by the advances in the field of digital and computational demography, with a particular focus on previous contributions using geotagged social media. Next, we describe the data and methods applied to generate our results. The following section first deals with the internal validation of the residency assumption comparing the resulting data with self-reported home locations from Twitter users. We then conduct an external validation using annual aggregates of county residents and available statistics of SC 2% accommodations tax collections. Next, we present the potential of the results with a case study based on the 2017 total solar eclipse and a low-high tourist season comparison. Finally, we conclude with a discussion about the potential applications of the approach for different disciplines and further work.

## 2. Related work

### *2.1. Measuring population movements*

Demography is a highly data-driven discipline that has sustained a significant share of its advancements in data collection efforts. Other disciplines with interest in studying population movements such as geography or tourism also share this concern about the availability of data to fuel their investigations. Many scholars have voiced the necessity of measuring population movements in a consistent manner considering all their dimensions, for which data quality is vital (Skeldon, 2012; Bell et al., 2015). Monitoring populations is critical to evaluate demographic trends, and researchers have traditionally relied on a set of data sources that includes vital statistics, censuses/registries, and population-based surveys (Mallick and Vogt 2014; Fussell et al. 2017). However, Coleman (2013) pointed out that the distinction between these sources is becoming fuzzier and that hybrid approaches tend to be employed.

Demographers interested in large-scale population movements rely on censuses and registries study these phenomena. Censuses and registries are systematic data-collection efforts carried out by official entities to record information about a given population. Censuses are internationally accepted, and the United Nations issues standards and methods to assist national statistical authorities in their compilation (United Nations 2008). The main limitation of these approaches is its cross-sectional temporal coverage, with a considerable time lag between collection periods, which is insufficient for most migration research purposes (Fussell et al., 2014b). Many countries only carry out censuses on a five- or ten- year basis. Some of the countries are systematic with these data-collection procedures and undertake the census campaign on fixed years, therefore offering an inventory of the population of a territory on a regular basis. However, this is not the case for all countries, and many have abandoned or not initiated a regular and universal data-collection strategy (Bell et al. 2015), which is often related to the considerable human and economic resources needed to carry out these initiatives (Fussell et al. 2014b). Although demographers have used censuses to study migration, they are not explicitly designed for this purpose and can therefore only include a limited number of questions about this matter. Bell et al. (2015) discussed the different mechanisms used to infer migration/mobility data from censuses (i.e., lifetime migration, migration over a fixed interval, or place of the last residence), and stresses the great differences among countries, which further limits the comparability of the data.

Nationwide surveys are another widely used data-collection method for internal migration and tourist movements. One of the main advantages over censuses or registries is that it can provide information with more temporal periodicity (more frequent data). In addition, national surveys can be tailored to reflect on large-scaled migration, daily mobility, or tourist behaviors, unlike censuses and registries whose focus is not on these issues. This flexibility that surveys allow, in addition to being less expensive and involving considerably fewer human resources than censuses and registries, has enabled this data-collection method to substitute universal data-collection strategies (i.e. censuses and registries) in some countries (Franklin and Plane, 2006). These surveys are normally available on a yearly basis, which is more amenable for migration studies, although it still does not suffice to comprehensively study short-term movements such as weekly/seasonal mobility patterns or special events such as emergency-induced population movements. One of the trade-offs of nationwide surveys (e.g. American Community Survey (ACS)) in comparison with census data is the loss of geographic detail (Bell et al., 2015). Annual surveys are not universal, which further generates problems of representativeness, particularly in small communities.

None of the traditional methods to keep track of population movements has shown to be completely adequate to study such dynamic and complex processes in a comparable and replicable manner, particularly in a context of increased spatial mobility. Thus, many scholars insist on incorporating innovative data collection methods that can complement conventional approaches and offer a more dynamic estimate of human spatial behavior (Willekens et al. 2016)

## 2.2. The digital geospatial shadow

According to multiple researchers, the study of population movements and spatial behavior has entered a new era or a new data paradigm (Isaacson and Shoval, 2006; Billari and Zagheni, 2017; Cesare et al., 2018). These scholars refer to a new phase where the study of population movements can be much more precisely and dynamically measured using geospatial digital trace data. This new area of study has grown considerably in the previous years, and a new field called digital and computational demography has emerged. Within this growing body of literature, many contributions have been made using different data sources and data-collection approaches. The final output of these investigations has improved the knowledge of population processes, such as fertility and mortality (Tamgno et al. 2013), migration (Zagheni and Weber, 2012), or emergency-induced population movements (Martín et al., 2019).

Digital and computational demography is based on the use of Big Data to track populations. Within the term of Big Data, we find multiple types of data. The one that holds the greatest potential for the field is geospatial digital trace data. These, also known as passive citizen sensor data, are generated by individuals in their daily digital activity. Passive citizen-sensor data can be used for many purposes, research being just one of them. These digital shadows, unintentionally left behind by the individuals, can be used to determine the spatial behavior of those who generated them, as some carry information about the physical location where the data record was created. The possibilities of these data go beyond its application in demography (e.g., Zagheni et al., 2014) and other fields or disciplines have also leveraged them. This is the case of transportation (Jurdak et al., 2015), public health (Wesolowski et al. 2012), sociology (Amini et al. 2014) or natural hazards (Martín et al., 2017; Li et al., 2018, Huang et al., 2019). Some of the characteristics that these researchers seek in passive citizen sensor data is its immediacy, as its collection and exploitation can be done close to real time, its wide coverage, and its reduced cost (Spyratos et al., 2018).

Among the different sources of passive citizen sensor data, mobile phone call detail records (CDR) has been the one with the largest number of applications, although data accessibility is a severe limitation, as these data are privately owned by corporations that are reluctant to freely share these data for research purposes. Although there have been a few initiatives to provide researchers with CDR data (e.g., Orange Telecom Data for Development Change or the Turk Telekom Data for Refugees challenge), data share agreements between phone carriers and researchers are scarce. Also, scholars have voiced concerns about ethical and privacy issues, in particular with vulnerable populations (Taylor, 2016). Despite these pitfalls, CDR is considered one of the richest passive citizen sensor data, as the penetration of cell phones in today's society is very high (Turner, 2020), and they provide high periodicity data (continuous records rather than sporadic or episodic). The application of CDR in short-term mobility and human spatial behavior studies has been extensive. For instance, Alexander et al. (2015) processed mobile phone data from millions of anonymized users to estimate daily origin-destination trips by purpose and time of day, comparing the results with local and national surveys. Also in relation to regular mobility patterns but focusing on their association with the transmission of infectious diseases, Wesolowski et al. (2014) leveraged a large dataset of CDR in fifteen West African countries and managed to connect these mobility patterns with the propagation of the Ebola virus. In the post-disaster context of the Haiti 2010 Earthquake and the subsequent cholera outbreak, Bengtsson et al. (2011) tracked 138,560 SIM cards to produce detailed reports of the population movements associated with the disaster. Long-term analyses using CDR are not common, one of the few studies in the literature was conducted by Blumenstock (2012), who investigated a four-year dataset from 1.5 million Rwandans and revealed patterns of temporary and circular migration previously unknown.

Due to the great difficulty of accessing mobile phone call records, many researchers have resorted to other sources of digital shadows to track populations. Virtually any kind of online platforms has potential to be used for the study of short-term or long-term mobility or migration. For instance, State et al. (2013) used logins to Yahoo! to estimate short and medium-term migration flows. In a similar study, Zagheni and Weber (2012) explored Yahoo! e-mails to identify age and gender-specific patterns of migration. Social media data have been largely utilized in many social science research purposes, including as a proxy to determine the spatial behavior of people. Social media is considered as the richest supplier of data due to the easier accessibility of its content (Stock 2018). Some of the platforms explored by researchers include video conference applications, such as Google+, used by Messias et al. (2016) to study migration clusters, thus helping to develop new theories of international migration that could not be tested with traditional approaches. Another video conference application explored is Skype. Kikas et al. (2015)

presented the results of a study of cross-country migration as observed via login events in the Skype network. Others have leveraged LinkedIn to focus on work-related population movements (State et al., 2014; Barslund and Busse, 2016). Researchers have also utilized some of the social media leading platforms such as Facebook or Twitter for their research purposes. Working around the limitation of Facebook's data sharing policy —much more restrictive than Twitter's— Zagheni et al. (2017) proposed an approach that leveraged Facebook's advertisement platform (which is freely available) to estimate the stock of international migrants in the US. In a more recent study, Alexander et al. (2019) investigated the impact of Hurricane Maria on emigration from Puerto Rico also using Facebook advertising platform (Ads Manager). Due to Twitter's open data sharing policy, this is the social media source of data that has garnered more attention from researchers, including migration and mobility scholars. For instance, Hawelka et al. (2014) explored tweets to estimate the magnitude of international travelers by country of residence in 2012, identifying patterns of global mobility. Martín et al. (2017) and Jiang et al. (2019) investigated the evacuation of Hurricane Matthew in the southeast of the US also using geotagged tweets.

Social media data, and Twitter data in particular, offer immediacy (i.e., data available almost in real-time) and adequate spatiotemporal coverage for many mobility and migration analyses. Its reduced cost —mainly only computational— and the large number of users in many countries are also seen as great advantages for research and management. Table 1 shows the twenty countries with the most Twitter users and their Twitter penetration rate. As we can see, the penetration rate is highly variable across countries. However, relatively small percentages mean millions of users and a potential sample size much larger than those obtained with traditional methods such as surveys. At this moment, there is no systematic approach that provides the research community and practitioners with ready-to-use (near) real time information about the daily number of users active in a region and their origin (resident of the area or non-resident). With this contribution and following research, we intend to fill this gap and offer a valuable dataset with the capacity to further the knowledge about different types of mobility and migration. Thus, this new proxy has the potential to introduce a dynamic component in the estimation of both resident and non-resident stocks, a much-needed information in several aspects such as research, tourism, or emergency management.

Table 1. Countries with the most Twitter users in 2020 and penetration rate. Source: Statista (2020)

| Countries | Total (million users) | Penetration rate |
|---|---|---|
| United States | 62.55 | 18.9% |
| Japan | 49.1 | 38.8% |
| India | 17 | 1.3% |
| Brazil | 15.7 | 7.5% |
| United Kingdom | 15.25 | 22.9% |
| Turkey | 12.7 | 15.5% |
| Saudi Arabia | 12 | 35.6% |
| Indonesia | 11.2 | 4.2% |
| Mexico | 10.4 | 8.2% |
| Philippines | 7.75 | 7.3% |
| Thailand | 7.15 | 10.3% |
| France | 7.1 | 10.6% |
| Spain | 7.1 | 15.1% |
| Canada | 5.8 | 15.4% |
| Argentina | 5.05 | 11.4% |
| South Korea | 4.7 | 9.1% |
| Germany | 4.2 | 5.1% |
| Egypt | 3.55 | 3.6% |
| Colombia | 3.35 | 6.7% |
| Malaysia | 3.1 | 9.8% |

## 3. Methodology

### 3.1. Data collection and processing

Geotagged Twitter data were collected using the Twitter Streaming Application Programming Interface (API) with a bounding box covering the whole world. It should be noted that Twitter Streaming API allows access to about one percent of the tweet streams. The tweets were cleaned and preprocessed into CSV (comma-separated values) files storing in a Hadoop-based big data computing cluster with customized spatiotemporal indexing. Hadoop Hive, Impala, and ESRI tools for Hadoop are used to conduct spatial analyses and query analytics.

The literature provides several methods to classify users into local (resident) users and non-resident users. Some have used density-based spatial clustering of applications with noise (DBSCAN) (Ester et al., 1996) to retrieve the user's home location (Huang and Xiao, 2015), which is the most precise method to obtain the home location at the finest possible spatial resolution when the geolocation is based on a pair of coordinates (latitude and longitude). However, the scope of our research is focused on the county level, which includes tweets with a spatial resolution finer than county (e.g., a city name) but that is not necessarily pinpointed with a pair of coordinates. In addition, DBSCAN implies considerable computation challenges considering the extent of our study, daily estimates for all counties in SC (and all counties in the U.S. as our next step). This precludes us from using this method and recommends other approaches. Others have based their home-location classification in the study of the center of mass (Hawelka et al., 2014) or the median center (Martín et al., 2017), which is in accordance to Jurdak et al. (2015) findings about the preference of people to tweet around half of the times from their most popular location (home location). Following this idea, another method to identify residency of users has involved weighting the most frequently visited (WMFV) cluster (Lin and Cromley, 2017) or just selecting their most repeated location (the mode in numerical terms) (McNeill et al., 2016)Different scenarios and conditions have been applied to this latter approach, including n-day time intervals (Jiang et al., 2018; Hu et al., 2019) or night-time tweets (Koylu, 2018). Unfortunately, these investigations often do not report a measure of the goodness of their home-location assumptions (validation), for which selecting the most appropriate method remains conditional to the objectives of the particular research project. In our case, the goal of this piece of research of presenting an automated classification of resident/non-resident Twitter users, and its final application to the whole US, demands a precise distinction of the home location, yet a friendly approach from the computational standpoint that is able to generate the results in near-real time.

For this reason, we adopted the most repeated location among the tweets sent by each individual user during the year of the study period (January 1$^{st}$, 2017 - December 31$^{st}$, 2017). In addition, we tested n-day time interval but the small change in the residency assumptions (between 2% and 5%) led us to use the most simplistic approach based on just the most repeated location. The workflow to obtain the total Twitter residents and non-residents per county per day is presented below and synthesized in Figure 1 for three hypothetical types of users for Jan. 1$^{st}$, 2017 in Richland County, SC.

1. For each day, for each of the 46 counties of SC, obtain a list of distinct (unique) users active in that county.
2. Retrieve all tweets available in our repository from each of the active users from the year of the study period. We used an entire year to eliminate the effect of seasonal variations had we chose a shorter timeframe.
3. Assign a distinctive code of the US county, or country if outside the US, from where each of the tweets was sent.
4. Compute the most repeated tweeting location for each user during the year of the study period.
5. Classify those whose most repeated tweeting location falls within the analyzed county as *residents*. If a user's most repeated tweeting location is outside of the county, the program classifies this user as *non-resident*. A third category, *unknown*, is reserved for those users whose most repeated tweeting location cannot be calculated (when having less than two tweets in the whole year, or more than one county as the most repeated tweeting location).

A program was developed to iteratively perform the tasks in the workflow, running on the Hadoop cluster consisting of thirteen computer servers. It took about 20 minutes to automatically compute the daily number of users (resident, non-resident, and unknown) for each county of SC for the whole year of 2017.

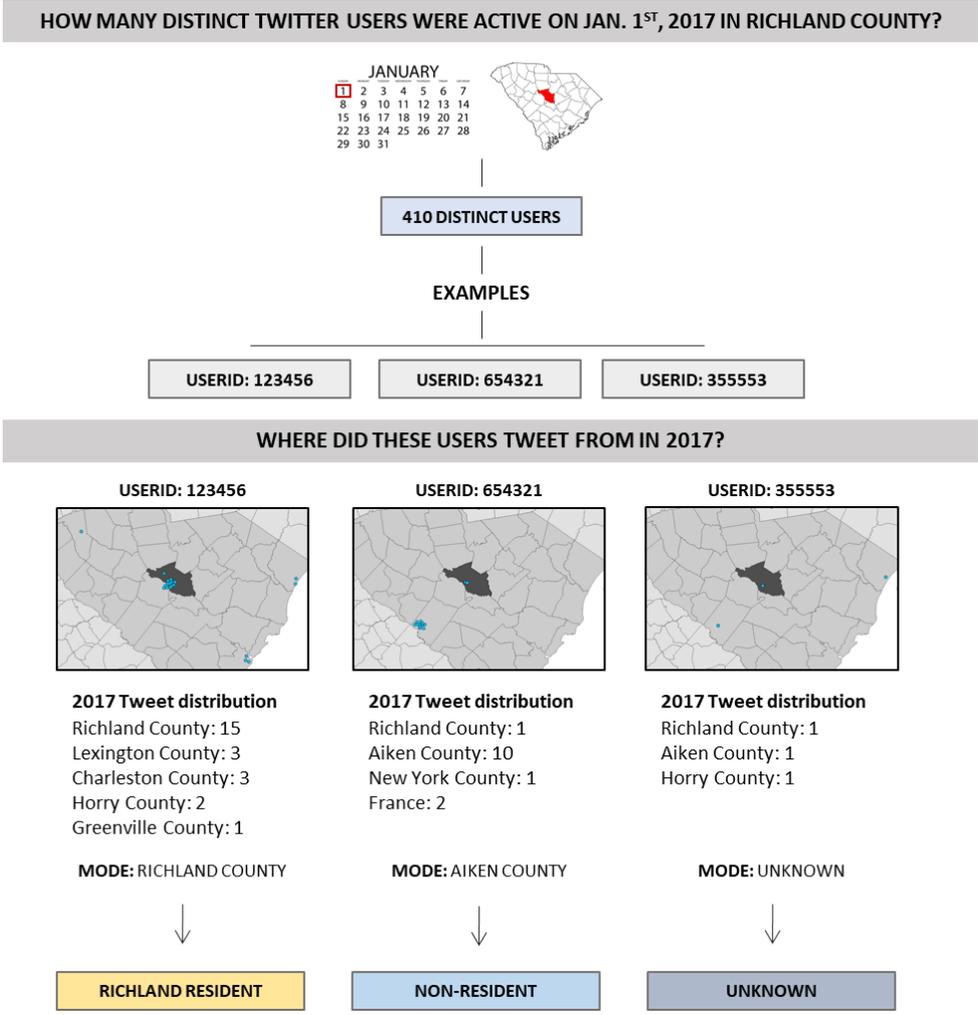

Figure 1. Example of workflow applied to three hypothetical types of users for January 1st, 2017 in Richland County, SC.

## 3.2. Validation analysis

To evaluate the performance of our automated classification we designed two validation analyses— internal and external. In addition, we use the 2017 Solar Eclipse of August 21st as operational validation to show the potential of the method.

### 3.2.1. Internal validation: residency assumption

The internal validation is designed based on the self-reported home location of a sample of Twitter users. Using self-reported home location is a common strategy to determine the residence of the social media user. For instance, according to Zagheni et al. (2018), this is the approach followed by Facebook Adverts Manager to defining expats (people whose original country is different from the current country). The process is summarized in the following steps:

1. Extraction of a representative sample of users. We calculated the total number of distinct active Twitter users (who tweeted at least once) in 2017 in SC. 190,608 distinct users were identified. We randomly selected a significant sample size of these population of Twitter users with a confidence level of 99% and a margin of error of 5%. The sample size was therefore fixed in 661 Twitter users with self-reported home location.
2. Manual assessment of the self-reported home locations of the sampled users to eliminate obvious fake locations (e.g. "Neverland", "Waffle House", "Mars", or "Moon") or not geographically detailed enough descriptions to estimate the home location at the county level (e.g., "South Carolina", or "United States"). 18.2% of the initial sample reported fake locations while 22.0% did not provide enough geographical detail. We randomly substituted these users with other users whose self-reported location was not fake and accurate enough until obtaining 661 users.
3. Manual attribution of a county from the self-reported location for each validation user. For instance, if a user self-reported his home location in Columbia, SC, we assigned Richland County.
4. Comparison of automatically estimated home location based on the most repeated tweeting county (mode) and the validation sample (failure/success analysis).
5. Individual assessment of the misidentifications.

### 3.2.2. External validation of final estimates

The second validation assessment aims to relate the obtained results with the most accurate and spatiotemporally detailed data about the number of residents and tourist activity. We used 5-year ACS estimates to relate the annual number of distinctive users per county with county resident estimates for 2017. The ACS is an annual survey of the U.S. Census Bureau designed to supplement the decennial census. There are no official statistics about the daily or seasonal variations of resident populations. Therefore, our daily data had to be aggregated to annual totals in order to be compared with ACS data. For tourism statistics, the only data available at the county level for SC is the monthly state accommodation tax (SCPRT, 2019), which is a mandatory 2% charge applied to all accommodations statewide. Accommodations are defined as "the rental or charges for any rooms, campground spaces, lodgings, or sleeping accommodations furnished to transients by any hotel, inn, tourist court, tourist camp, motel, campground, residence, or any place in which rooms, lodgings, or sleeping accommodations are furnished to transients for a consideration"(SCPRT, 2019). Again, there is no official data available at the daily level, and our daily estimations had to be aggregated, in this case to monthly data. We are aware that the state accommodation tax is just a partial estimate of the non-resident movements, as it does not register trips that do not involve accommodation or those who spend the night at second homes or relative/friends' homes.

### 4. Validation results

### 4.1. Internal validation: residency assumption

One of the objectives of the internal validation was to determine whether a threshold about the minimum number of tweets from users in the previous year was needed in order to accurately identify their home location. We report in Table 2 the percentages of successful identification of home residency by our method (most repeated tweeting county or mode) in comparison to the user's self-reported home location. As we can see in the table, once assigned the home-location category "unknown" to those whose most repeated tweeting county location could not be calculated, the percentage of success (between 77 and 78%) is almost independent from the number of tweets per user per year for users with less than 200 tweets. Only for users with 200 tweets or more in the year (26.2% of the total users) the home-location identification slightly improves by roughly 5%. For this reason, we decided not to apply any threshold in our method (*All users*).

Table 2. Percentages of successful home-location identification for different thresholds

|  | **Users** | **% Total** | **% Successful** |
|---|---|---|---|
| Cannot compute mode | 28 | 4.24% | / |
| All users* | 633 | 100.0% | 77.3% |
| Users 3 tweets or more* | 631 | 99.7% | 77.3% |
| Users 5 tweets or more* | 616 | 97.3% | 77.6% |
| Users 10 tweets or more* | 582 | 91.9% | 77.8% |
| Users 15 tweets or more* | 547 | 86.4% | 77.1% |
| Users 20 tweets or more* | 515 | 81.4% | 77.7% |
| Users 25 tweets or more* | 483 | 76.3% | 78.1% |
| Users 30 tweets or more* | 466 | 73.6% | 78.1% |
| Users 50 tweets or more* | 387 | 61.1% | 78.3% |
| Users 100 tweets or more* | 274 | 43.3% | 78.5% |
| Users 200 tweets or more* | 166 | 26.2% | 82.5% |
| Users 500 tweets or more* | 58 | 9.2% | 82.8% |
| Users 1000 tweets or more* | 18 | 2.8% | 83.3% |

*After removing those whose mode could not be computed

We believe 77-78% of successful home-location identification is sufficiently accurate for an automated method, more even so after individually analyzing the misidentifications. Looking at the home-location identification failures of users in the category "All users" from Table 1 (no threshold applied), we found out that 37 of the misidentifications (26%) could be related to the user identifying himself as a resident of a multicounty metropolitan area or with Twitter inaccurately placing a geotag (place) in a county that does not belong. Table 3 shows these issues, which relate to the validation process, and that might make the percentage of successful home-identification to be lower. In the case of Fulton and Gwinnet (GA), the user identified himself as an Atlanta (GA) resident; however, most of its tweeting activity falls within Gwinnett County, part of the metropolitan area of Atlanta. This problem was also common in Charleston (SC) and Columbia (SC) metropolitan areas. For instance, our method assigned home location in Berkeley or Dorchester (in towns such as Summerville, Goosecreek, or Hanahan) to users who self-reported Charleston as home location. This is particularly relevant in the SC capital city, Columbia, where the Congaree River divides the metropolitan area into two counties (Richland and Lexington) and multiple cities (Columbia, West Columbia, Cayce, or Lexington). Thus, 10.4% of the home-location misidentifications occurred in this area, where the user self-reported to be a resident of Columbia (Richland County) but its tweeting behavior showed his home residence in places like West Columbia, Cayce or Lexington (Lexington County). Similarly, this issue was also observed in Greenville and Pickens counties, where the users would self-report Greenville as home-location but their tweeting behavior would show that their home residency (most repeated tweeting location) is in the metropolitan area (partly pertaining to Pickens County). Finally, we also identified how Twitter wrongly placed the town of Fort Mill (SC) in Lancaster County, when it is located in York County. This error transfers into an automated home-location identification and is external to our identification approach. Speculating that the 37 misidentifications (32 from metropolitan areas issues and 5 because of Twitter geolocation of Fort Mill) were actually successful home-location identifications, the percentage of success reported in Table 1 would increase from 77.3% to 83.1%.

Table 3. Issues identified during the analysis of home-location misidentifications

| Affected counties | Users | % Total |
|---|---|---|
| Fulton (GA) / Gwinnett (GA) | 1 | 0.7% |
| Charleston / Dorchester / Berkeley | 11 | 7.6% |
| Richland / Lexington | 15 | 10.4% |
| Pickens / Greenville | 5 | 3.5% |
| York / Lancaster | 5 | 3.5% |

**4.2. External validation of final estimates**

Here we present the relationship of our Twitter classification of residents and non-residents with the most spatiotemporally detailed official statistics. Figure 2 shows the association between the estimated total county population and the daily average of total active Twitter residents per county. As we can see, the annual average of daily Twitter residents is highly correlated with the ACS estimates, and 85 percent of the variability in the data is explained by a straight line through the observations. We only identified one big outlier corresponding with Horry County (Myrtle Beach area) for which we have no explanation at this time. The rest of the observations are close to the line confirming that more Twitter residents are found in more populated counties.

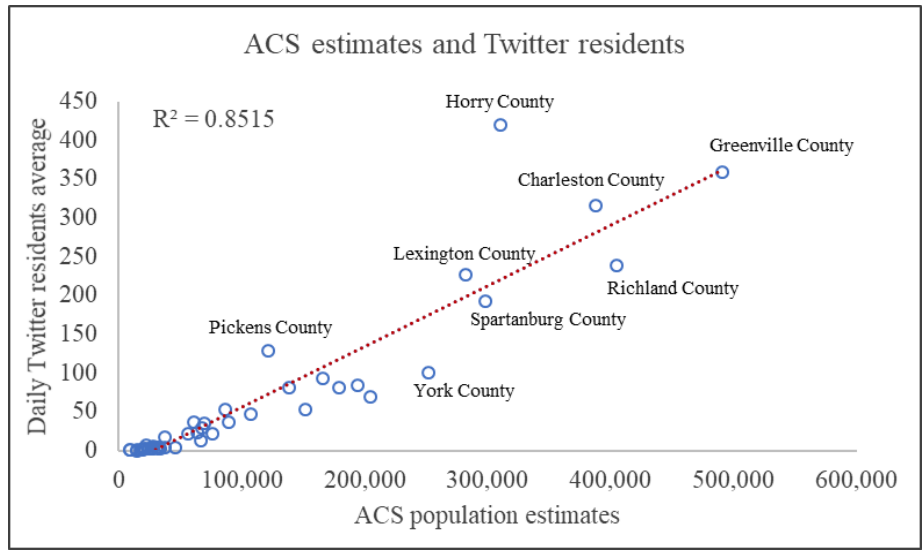

Figure 2. Scatter plot of 5-year ACS county population estimates and the daily average of total active Twitter residents per county (46 counties in total).

Figure 3 relates the monthly aggregated observations of non-resident Twitter users with SC Statewide 2% Accommodations Tax Collections. In the monthly log-log plots we can observe there is a strong positive linear relationship between the two variables. The $R^2$ values further reveal a closer relationship during the months of highest accommodation occupancy (summer months), with around 90% of the variance of June and July explained by a straight line. We believe the lower values during the rest of the months (still with strong relationships with $R^2$ over 0.6) are explained by the difference in the travel behavior. During the low (touristic) season, from September to February, travel tend to be more locally based and shorter duration, not reflecting as much on accommodation occupancy as much of this travel are day-long trips or local SC population with second homes in touristy areas (e.g. Myrtle Beach). The SC Department of Parks, Recreation, and Tourism did not have available data of accommodation tax collections for November 2017.

Overall, we are aware that the nature of the official data used for the external validation (ACS annual resident estimates and monthly aggregates of accommodation tax collections) does not fully confirm the goodness and fit of our approach. While these data constraints hinder the assessment of a geotagged social media approach to provide a daily picture of the behavior of residents and non-residents at the county level, they reveal the necessity of research of this kind that offers innovative solutions to the paucity of data about population stocks and mobility in such a spatiotemporal detailed scale.

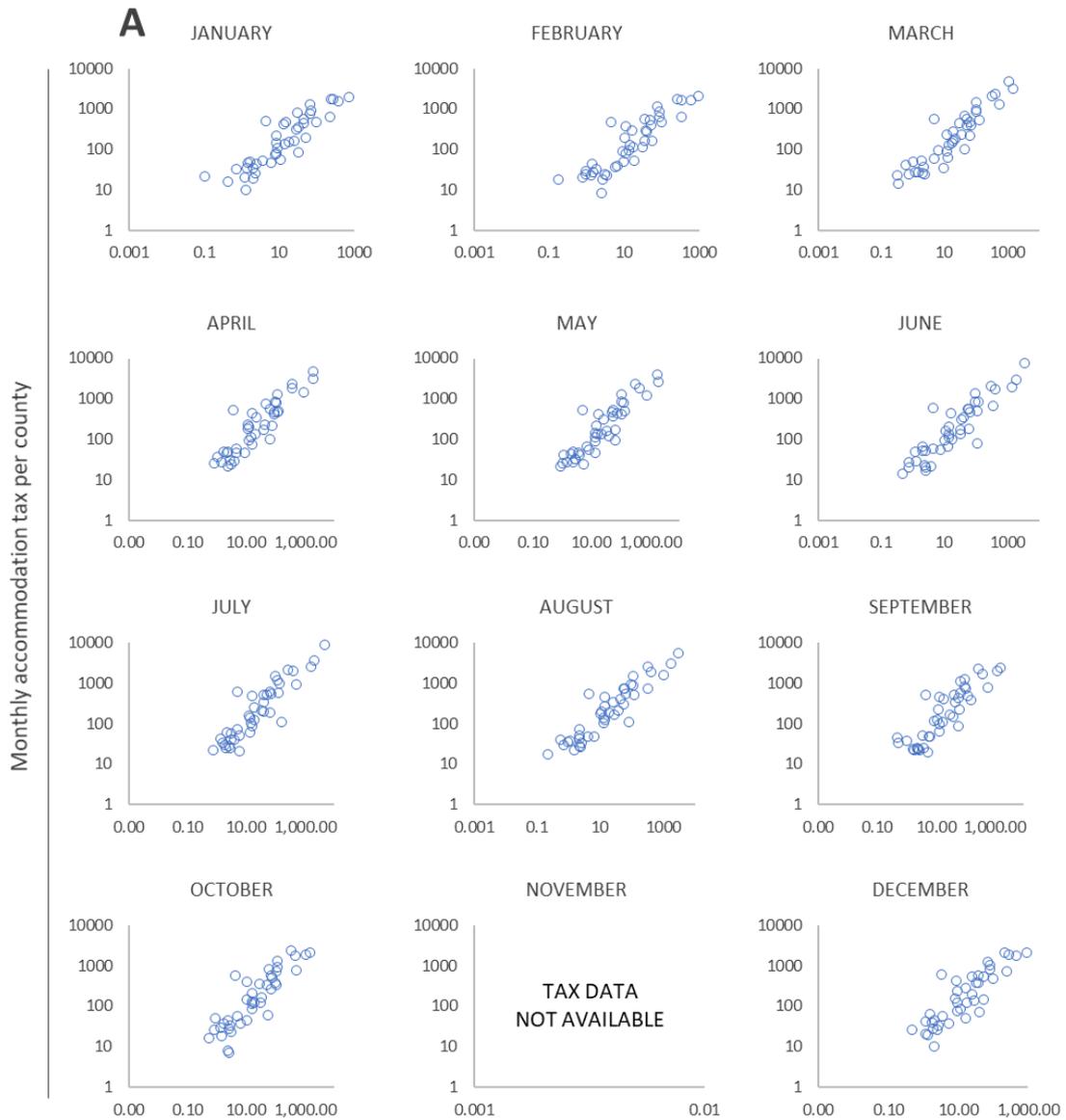

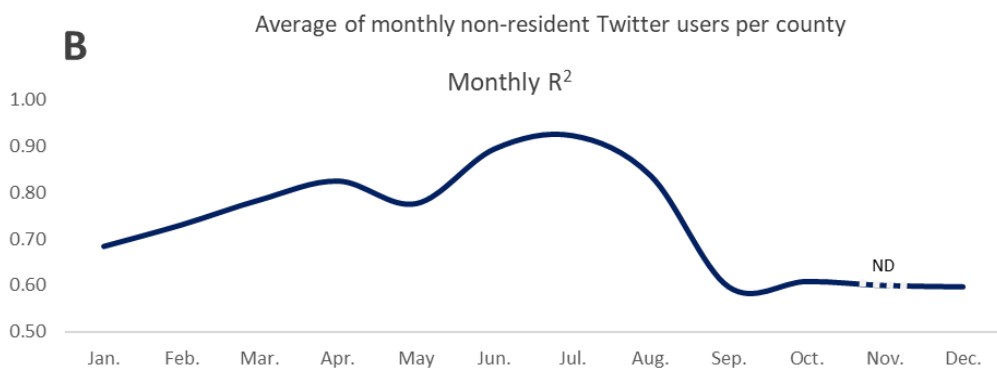

Figure 3. A) Log-log plots of the relationship between monthly accommodation tax per county and the average of monthly non-resident Twitter users per county. B) $R^2$ values of this relationship.

## 5. Case study: 2017 US Solar Eclipse and low-high tourist season comparison

The 2017 US Solar Eclipse offers a great opportunity to test the potential of our approach to detect, quantify, and compare the influx of Twitter non-residents to SC counties within the path of totality and to SC counties outside the path of totality (see Figure 3). On Monday August 21, 2017, a total solar eclipse moved across the continental US from the northwest (Oregon) to the southeast (SC). Millions of people traveled nationally and internationally to see the astronomic event. Some have pointed out this was the "greatest temporary mass migration of humans" in the US history, which put major pressure on infrastructures as near 200 million people lived in a one-day drive from the totality path, with SC being the closest location for near 95 million of them (Zeiler, 2017). Many cities and towns across the path of totality saw their population multiplied, many accommodations were 100% booked months in advance and many roads experienced bumper-to-bumper traffic during the previous weekend and on the Eclipse's day (Boyle, 2017). As we can see in Figure 3, the path of totality covered nearly half of SC in a 70-mile-wide swath. Thus, if our method is operatively valid, we can expect a major influx of Twitter non-residents in the SC counties under the totality path and a smaller increase in counties outside of it.

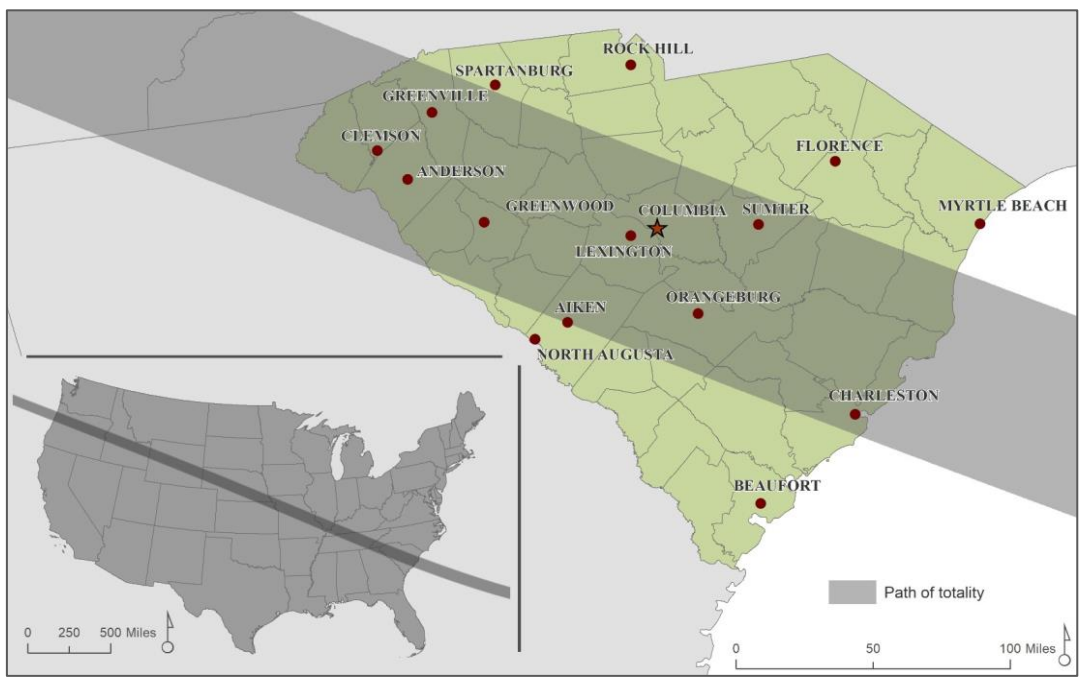

Figure 3. Path of totality in South Carolina.

The following video animation[1] shows two different (contrasted) scenarios of tourist visitations in SC: a low season situation during January 2017 and a high season situation during August 2017 (particularly the first half of the month). The video animation is only focused on these two months to reduce the length of the animation and simplify its interpretation. The mapping was developed to facilitate comparison along the year and across the different counties. For this reason, a 9-interval divergent legend (fewer Twitter non-residents than the average in brown shades and more Twitter non-residents than the average in green shades) was designed to represent the daily standard deviations from the mean of the daily percentage of Twitter non-residents (daily count of Twitter non-resident / annual cumulative total of Twitter non-residents). For visualization purposes and to incorporate the magnitude of actual number of Twitter non-residents per county, we randomly generated dots (1 dot = 3 Twitter non-residents) in the most densely populated census tracts of each county.

---

[1] https://youtu.be/O9qH4kw84MM

As we can see in the animation, January begins (particularly Jan. 1$^{st}$) with many counties experiencing positive standard deviations (green colors) as a consequence of the last days of the Christmas vacation period. However, this pattern soon dissipates and grey and brown shades dominate across SC during the whole month of January, only slightly altered during weekends. There were not many events that attracted non-residents to SC counties during January 2017, but we can identify two main occasions where we observe a major influx of non-residents in the northwest of the state (Oconee County, Pickens County, and Anderson County), both related to Clemson Football team. The first one, on January 9 and January 10, coincides with the 2017 College Football Playoff National Championship. Even though the game was played at Raymond James Stadium in Tampa (Florida), many Clemson fans gathered in these counties to watch the game with families and friends. The second event, a few days later, on January 14, coincides with the Clemson National Championship Parade and the championship celebration, which attracted a large number of fans to the area, as it can also be observed in Figure 5. August begins with a much different pattern where we can see green shades dominating the coastal counties of SC, particularly during weekends and especially in Horry County (one of the most touristic coastal destinations in the country). This high season period ends around mid-August and grey and brown shades dominate again throughout the state.

On Saturday, August 19 (Figure 4A) we can identify the first effects of the Solar Eclipse on the influxes of non-resident Twitter users, with several counties with +1 or +1.5 standard deviations from the county's year non-resident average. On August 20 (Figure 4B), the pattern becomes more obvious, particularly centered in the Columbia metropolitan area (Richland and Lexington counties) and the central coastal counties. On August 21 (Figure 4C), the day of the solar eclipse, most of the state recorded over 2 standard deviations above the mean in the daily percentages of non-residents, indicating that this day was exceptional for most of the counties, especially those in the path of totality. Green shades are still present on August 22, indicating that many visitors did not leave SC right after the solar eclipse. By August 23, the low season scenario returns and grey and brown shades are predominant.

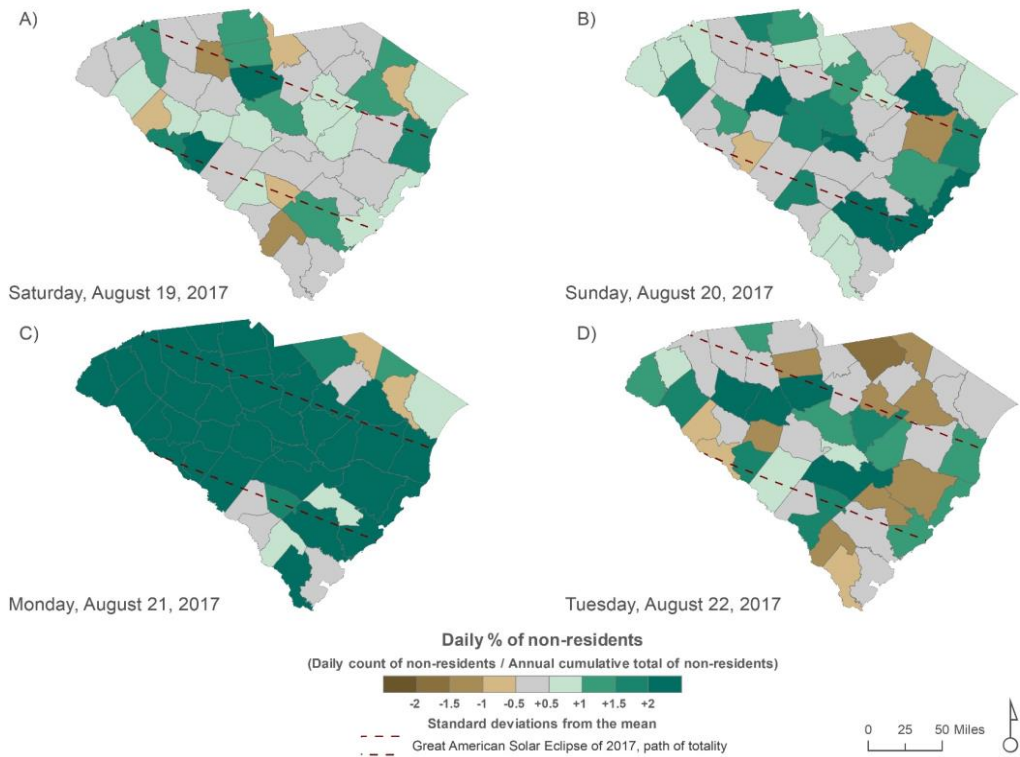

Figure 4. Standard deviations from the mean of daily percentage of Twitter non-residents from August 19 to August 22.

According to our data, counties in the totality path registered on average around 550% more Twitter non-residents in comparison with what could be considered normal for the third Monday of August. We estimated this normality by averaging the amount of Twitter non-resident of the following days: Monday, August 22, 2016; Monday, August 20, 2018; and Monday, August 19, 2019. Counties outside of the totality path registered 180% more Twitter non-resident than the normal values. Figure 5 shows the annual distribution for five counties (two outside the totality path and three inside of this totality path). Horry and Darlington counties, outside of the totality path, do not have a maximum on the day of the eclipse. Horry County shows a low-high season pattern with a maximum on July 4[th]. Darlington county has a major peak on Labor Day Weekend, coinciding with the annual celebration of the Bojangles' Southern 500 NASCAR race, which attracts to this small community tens of thousands of visitors during the whole weekend. Pickens County shows an interesting Twitter non-resident distribution across the year. Being in the totality path, we can observe a great influx of non-residents around August 21, but this is not the highest peak of the year. This county, home of the Clemson college football team, has a multipeak distribution coinciding with Clemson home games during the fall. Lexington County, in the central area of SC, does not stand out as a touristy location, but it did experience a major arrival of non-residents around the eclipse. In Charleston graphic, we can easily observe a serrated distribution across the year due to weekday-weekend duality (higher visitation during weekends) and a pronounced peak coinciding with the eclipse.

The results of this case study confirm that the presented approach is operatively valid to estimate the magnitude of non-resident visitations at the county and daily scales. This research not only validates the resident/non-resident classification approach employed but showcases the potential application of the method for real-time population estimation when combined with other sources of data such as surveys (Alexander et al. 2020)

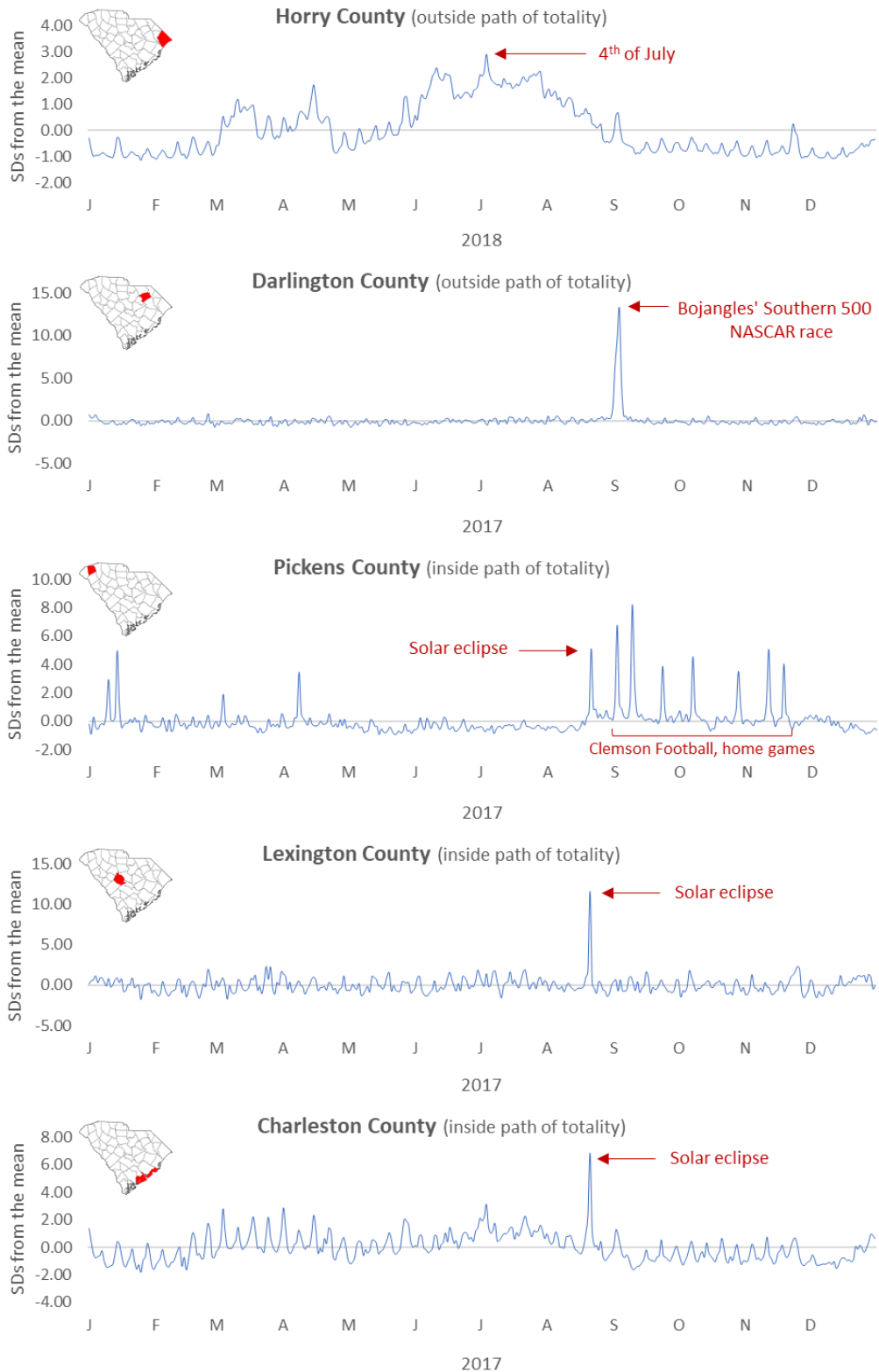

Figure 5. Annual distribution on Twitter non-residents (expressed in daily standard deviations from the mean) for five SC counties: outside of path of totality (Horry County and Darlington County) and inside of the path of totality (Pickens County, Lexington County, and Charleston County).

## 6. Conclusions and future work

The field of digital and computational demography is yet to fully harness the potential of digital trace data and produce population statistics with more suitable spatiotemporal resolutions for the study of population dynamics. This work intends to be a first step towards an automated Twitter census that can provide a more temporally granular picture of the local and non-local populations at the county level. Focusing on data from 2017 in SC, we designed a scalable (from the computational perspective) method that precisely determines —success rate around 80% according to internal validation based on self-reported home location— the residential location of a Twitter user based on their tweeting activity. An external validation confirmed that the yearly aggregated results of Twitter residents from our method correlate well with 5-year ACS county population estimates. In addition, the monthly aggregated Twitter non-residents found a strong positive relationship with a tourism indicator such as the monthly state accommodation tax at the county level. Altogether, this external validation helped to confirm the validity of the approach.

The case study of the 2017 US Solar Eclipse further confirmed the potential and validity of the approach as it added intuitive sense to the data resulting from our method. The method was able to estimate the magnitude of the extraordinary influx of Twitter users to SC, in particular to counties under the path of totality, by comparing it with normal expected visitation levels. The resulting data was also able to identify high/low visitation seasons, weekday/weekend visitation patterns, and detect other events, such as college football games or a NASCAR race. It is important to note that our approach did not account for potential increases of tweeting activity during special events. In other words, people might tend to tweet more frequently in particular events such as disasters, which could increase the identified resident and non-resident users in a county. This limitation will be addressed when we extend the analysis to the whole US/World and for more years than 2017 by offsetting/normalizing this effect using the total daily tweets sent in each county. In follow-up projects, we intend to make the information available in near real-time, which it would be of great help to follow dynamic processes such as evacuations or shelter-in place orders during emergencies. We believe this new dataset could have an impact in fields as varied as demography, geography, tourism, emergency management, and public health and create new opportunities for large scale mobility analyses, as it can introduce the dynamic component often lacking in population estimates. Even though this is just an initial step and more detailed analyses would be needed to blend Twitter with other sources such as surveys and SafeGraph data (https://www.safegraph.com) and model the intrinsic biases associated with social media data, geotagged tweets, as one of the few forms of freely available geospatial digital trace data, hold promise for the understanding of human spatial behavior under normal and extraordinary conditions such as the COVID-19 global crisis.